\documentclass{appolb}
\usepackage{graphicx}
\begin{document}
\title{Hadronic and rare B decays with the BaBar and Belle experiments%
\thanks{Presented at Cracow Epiphany Conference, 9-11 January 2012}%
}
\author{Xavier Prudent}
\address{Institut f\"ur Kern- und Teilchenphysik, Technische Universit\"at Dresden}
\\
%{Third Author of different affiliation
%}
%the Name(s) of other Author(s)
%\address{affiliation}
%}
\maketitle
\begin{abstract}
We review recent experimental results on $B_d$ and $B_s$ mesons decays by the BaBar and Belle expeiments. These include measurements
of the color-suppressed decays $\bar{B}^0\rightarrow D^{(*)0}h^0,h^0=\pi^0,\eta,\eta^{\prime},\omega$, observation of the baryonic decay $\bar{B}^0\rightarrow\Lambda_c^+\bar{\Lambda}K^-$, measurements of the charmless decays $B\rightarrow\eta h,h=\pi,K$, $B\rightarrow K\pi$, and observation of CP eigenstates in the $B_s$ decays: $B^0_s\rightarrow J/\psi f_0(980)$, $B^0_s\rightarrow J/\psi f_0(1370)$ and $B^0_s\rightarrow J/\psi\eta$. The theoretical
implications of these results will be considered.

\end{abstract}
\PACS{14.20.Mr}
  
\section{Introduction}

Given the large mass of the top quark, $B$ mesons are the only weakly decaying
mesons containing quarks of the third generation. Their decays are thus a unique
window on the Cabibbo-Kobayashi-Maskawa (CKM) matrix elements,
describing the couplings of the third generation of quarks to the lighter quarks. Hadronic $B$
mesons decays occur primarily through the Cabibbo favored $b\rightarrow c$ transition.
In the Standard Model these decays can also occur through Cabibbo suppressed $b\rightarrow u$ transitions or through one loop diagrams, such as penguin diagrams, which
involve a virtual $W^{\pm}$ boson and a heavy quark.
This proceeding reviews recent results~\cite{colsup}\cite{baryo}\cite{etah}\cite{Kpi}\cite{jpsif0}\cite{jpsieta} from the BaBar~\cite{babar} and Belle~\cite{belle} experiments
which took data during the past decade at the high luminosity $B-$factories PEP-II~\cite{pep} and KEKB~\cite{kek}.

\section{Color-suppressed decays $\bar{B}^0\rightarrow D^{(*)0}h^0,h^0=\pi^0,\eta,\eta^{\prime},\omega$}

In such decays, the effect of color suppression is obscured by the exchange of soft gluons (final state interactions),
 which enhance $W^{\pm}$ exchange diagrams. Previous measurements of the branching fractions of the color-suppressed decays 
$\bar{B}^0\rightarrow D^{(*)0}h^0$ invalidated the factorization model~\cite{colsup1}\cite{colsup2}\cite{colsup3}. However more precise measurements are needed to confirm that result and to constrain the different
QCD models: SCET (Soft Collinear Effective Theory) and pQCD (perturbative QCD). BaBar measured the branching fractions from
exclusive reconstruction using a data sample of $454\times 10^6 B\bar{B}$ pairs~\cite{colsup}, the measured values can be found in the Table~\ref{tab:colsupBF} compared to theoretical predictions. The values measured are higher by a factor of about three to five than the values predicted by factorization. The pQCD predictions are closer to experimental values but are globally higher, except for the $D^{(*)0}\pi^0$ modes. SCET~\cite{SCET_1}\cite{SCET_2}\cite{SCET_3} does not give prediction on the branching fractions themselves, but predicts that the ratios $BF(\bar{B}^0\rightarrow D^{*0}h^0)/BF(\bar{B}^0\rightarrow D^{0}h^0)$ are about equal to one for $h^0=\pi^0,\eta,\eta^{\prime}$. The ratios of branching fractions are given in Table~\ref{tab:colsupRatio} and are compatible with one. 
This SCET prediction holds only for the longitudinal component  $\bar{B}^0\rightarrow D^{(*)0}h^0$, in the case of $h^0=\omega$ nontrivial long-distance QCD interactions may increase the transverse
amplitude. The longitudinal fraction $f_L$ of $B$ decays to a pair of vector mesons is predicted to be one in the factorization description.
The longitudinal fraction of the decay $\bar{B}^0\rightarrow D^{(*)0}\omega$ was measured for the first time in the same data sample, yielding $f_L=(66.5\pm 4.7(\textrm{stat.})\pm 1.5 (\textrm{syst.}))\%$~\cite{colsup}, deviating thus significantly from the factorization's prediction. This reinforces the conclusion
drawn from the branching fraction measurements on the validity of factorisation in color-suppressed decays and supports expectations from SCET.

\begin{table}[htb]
\caption{\label{tab:colsupBF}
  Comparison of the measured branching fractions $BF$, with the predictions
  by factorization \cite{Chua,Neubert,Deandrea,Deandrea2}
  and pQCD~\cite{pQCD_1,pQCD_2}. The first quoted uncertainty is
  statistical and the second is systematic.}
\begin{center}
\begin{tabular}{lcccccc}
  \hline\hline 
$BF$  $(\times 10^{-4})$ & & This measurement  & & Factorization & & pQCD \\
\hline 
$\bar{B}^0\rightarrow D^0\pi^0$  & &  2.69 $\pm$ 0.09 $\pm$ 0.13 &  &0.58~\cite{Chua}; 0.70~\cite{Neubert} & & 2.3-2.6 \\
  $\bar{B}^0\rightarrow D^{*0}\pi^0$ & & 3.05 $\pm$ 0.14 $\pm$ 0.28 &  & 0.65~\cite{Chua}; 1.00~\cite{Neubert} & & 2.7-2.9 \\
  $\bar{B}^0\rightarrow D^0\eta$ & &  2.53 $\pm$ 0.09 $\pm$ 0.11   &  & 0.34~\cite{Chua}; 0.50~\cite{Neubert} & & 2.4-3.2\\
  $\bar{B}^0\rightarrow D^{*0}\eta$ & & 2.69 $\pm$ 0.14 $\pm$ 0.23 &  & 0.60~\cite{Neubert} & & 2.8-3.8 \\
  $\bar{B}^0\rightarrow D^0\omega$ &  & 2.57 $\pm$ 0.11 $\pm$ 0.14 &  & 0.66~\cite{Chua}; 0.70~\cite{Neubert} & & 5.0-5.6 \\
  $\bar{B}^0\rightarrow D^{*0}\omega$ & &  4.55 $\pm$ 0.24 $\pm$ 0.39 & &  1.70~\cite{Neubert} & & 4.9-5.8  \\
 $\bar{B}^0\rightarrow D^0\eta^{\prime}$ &  & 1.48 $\pm$ 0.13 $\pm$ 0.07 &  & 0.30-0.32~\cite{Deandrea2}; 1.70-3.30~\cite{Deandrea} & & 1.7-2.6  \\
 $\bar{B}^0\rightarrow D^{*0}\eta^{\prime}$ & &  1.48 $\pm$ 0.22 $\pm$ 0.13   & &  0.41-0.47~\cite{Deandrea} & & 2.0-3.2  \\
\hline\hline
\end{tabular}
\end{center}
\end{table}

\begin{table}[htb]
  \caption{\label{tab:colsupRatio}
    Ratios of branching fractions $BF(\bar{B}^0\rightarrow D^{*0}h^0)/BF(\bar{B}^0\rightarrow D^0 h^0)$.
    The first uncertainty is statistical, the second is systematic.}
  \begin{center}
    \begin{tabular}{lccc}
      \hline\hline
      $BF$ ratio & & & This measurement\\ \hline
      ${D^{*0}\pi^0}/{D^0\pi^0}$ &  && 1.14 $\pm$ 0.07 $\pm$ 0.08 \\
      ${D^{*0}\eta(\gamma\gamma)}/{D^0\eta(\gamma\gamma)}$ & & &  1.09 $\pm$ 0.09 $\pm$ 0.08 \\
      ${D^{*0}\eta(\pi\pi\pi^0)}/{D^0\eta(\pi\pi\pi^0)}$ & & & 0.87 $\pm$ 0.12 $\pm$ 0.05 \\
      ${D^{*0}\eta}/{D^0\eta}$ (Combined) &  & &  1.03 $\pm$ 0.07 $\pm$ 0.07 \\
      ${D^{*0}\omega}/{D^0\omega}$ &  & &  1.80 $\pm$ 0.13 $\pm$ 0.13 \\
      ${D^{*0}\eta^{\prime}(\pi\pi\eta)}/{D^0\eta^{\prime}(\pi\pi\eta)}$ & & &  1.03 $\pm$ 0.22 $\pm$ 0.07 \\
      ${D^{*0}\eta^{\prime}(\rho^0\gamma)}/{D^0\eta^{\prime}(\rho^0\gamma)}$ & & & 1.06 $\pm$ 0.38 $\pm$ 0.09 \\
      ${D^{*0}\eta^{\prime}}/{D^0\eta^{\prime}}$ (Combined) & & &  1.04 $\pm$ 0.19 $\pm$ 0.07 \\
%      \hline \\
%${D^0\eta^{\prime}}/{D^0\eta}$ &  & &  0.54 $\pm$ 0.07 $\pm$ 0.01 \\
%${D^{*0}\eta^{\prime}}/{D^{*0}\eta}$ & & &  0.61 $\pm$ 0.14 $\pm$ 0.02 \\
\hline\hline
\end{tabular}
\end{center}
\end{table}

%%%%%%%%%%%%%%%%%%%%%%%%%%%%%%%%%%%%%%%%%%%%%%%%%%%%%%%%%%%%%%%%%%%%%%%%%%%%%%%%%%%%%%%%%%%%%%%%%%%%%%%%%%%%%
%%%%%%%%%%%%%%%%%%%%%%%%%%%%%%%%%%%%%%%%%%%%%%%%%%%%%%%%%%%%%%%%%%%%%%%%%%%%%%%%%%%%%%%%%%%%%%%%%%%%%%%%%%%%%

\section{Baryonic decay $\bar{B}^0\rightarrow\Lambda_c^+\bar{\Lambda}K^-$}

\begin{figure}[htb]
\begin{center}
\includegraphics[width=8cm]{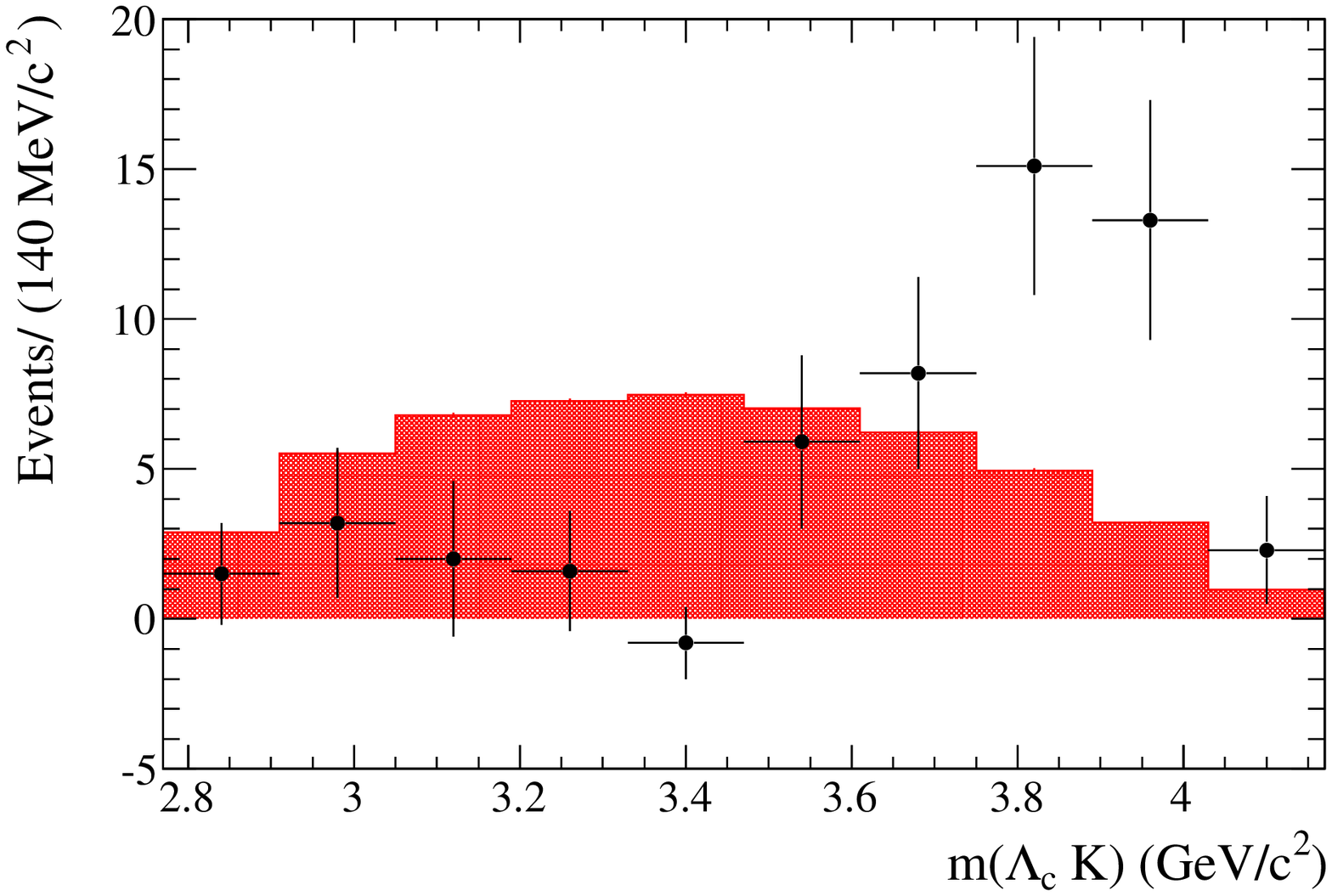}\\
\includegraphics[width=8cm]{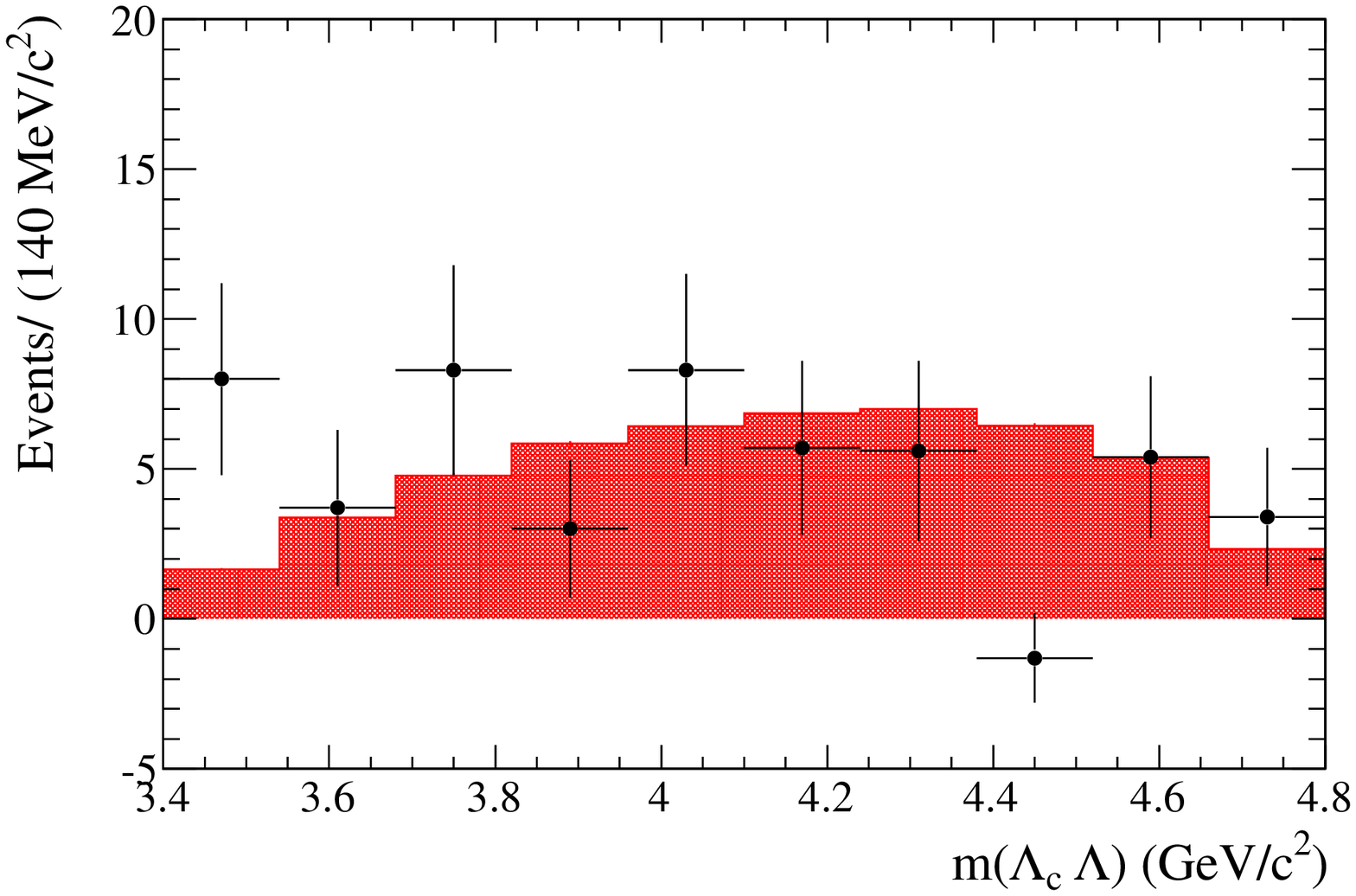}\\
\includegraphics[width=8cm]{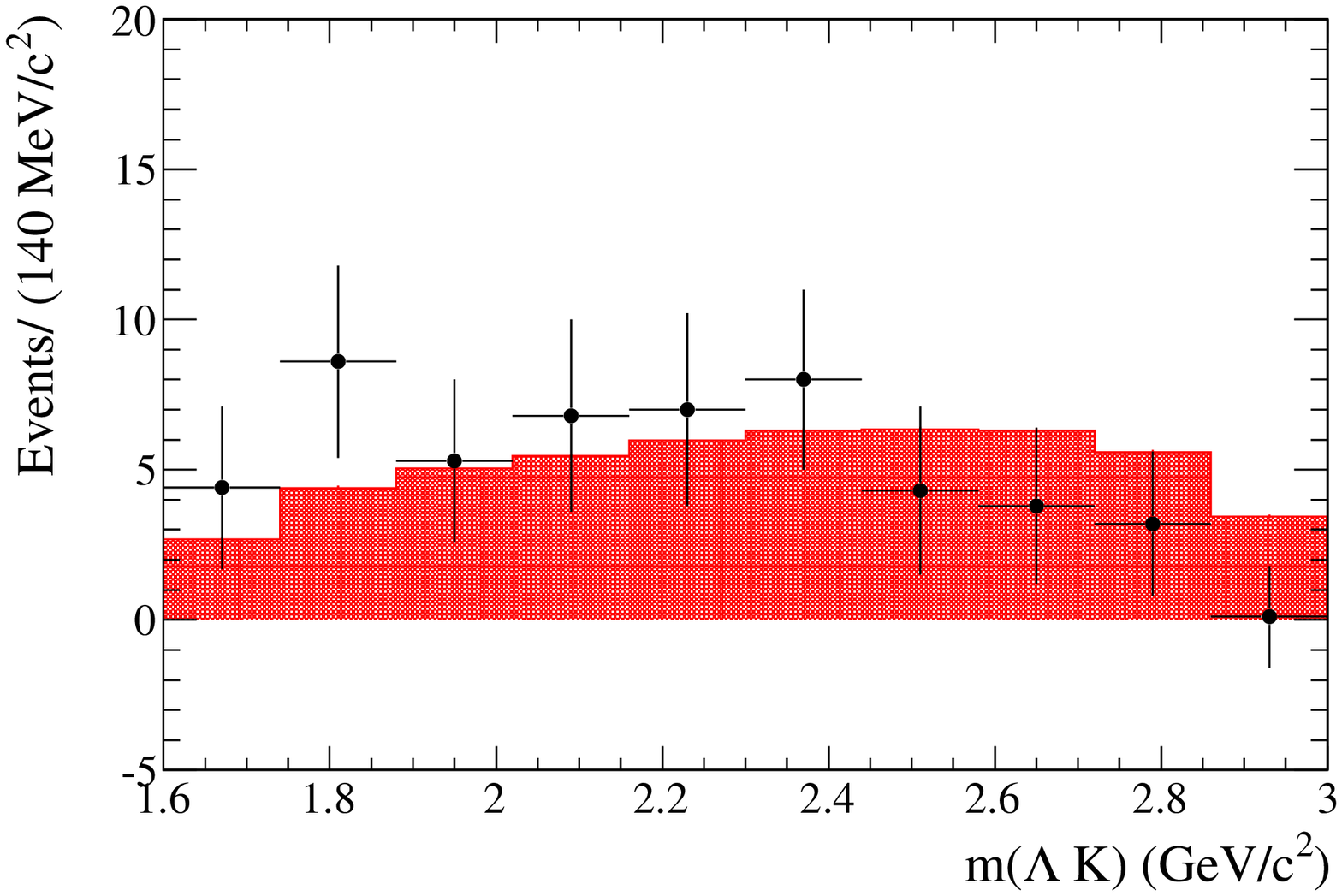}
\caption{Background-substracted distributions of the invariant masses $m(\Lambda_c K)$, $m(\Lambda_c\Lambda)$ and $m(\Lambda_ K)$ in data (points) and simulated Monte Carlo non-resonant signal sample (full histogram)}
\label{fig:baryoMasses}
\end{center}
\end{figure}

Baryonic decays account for $(6.8\pm 0.6)\%$~\cite{PDG} of all $B$ mesons decays, however little is know about these processes.
The reconstruction of exclusive final states allow to compare decay rates, and hence to increase our understanding of the fragmentation of $B$ mesons into hadrons.
The first measurement of the decay channel $\bar{B}^0\rightarrow\Lambda_c^+\bar{\Lambda}K^-$ is reported here~\cite{baryo}, using the full BaBar $\Upsilon(4S)$ sample, thus $471\times 10^6 B\bar{B}$ pairs. The background-substracted distributions of the invariant masses $m(\Lambda_c K)$, $m(\Lambda_c\Lambda)$ and $m(\Lambda_ K)$ are given in the Fig.~\ref{fig:baryoMasses}. A resonant structure is observed above $3.5~GeV/c^2$ in $m(\Lambda_c K)$, while no threshold enhancement is observed in $m(\Lambda_c\Lambda)$, in contrary to other three-body baryonic $B$ decays~\cite{ichep2008}. The branching fraction is measured after rescaling the simulated efficiency to the data distribution, yielding: $BF(\bar{B}^0\rightarrow\Lambda_c^+\bar{\Lambda}K^-)=(3.8\pm 0.8(\textrm{stat.})\pm 0.2(\textrm{syst.})\pm 1.0(\Lambda_c^+))\times 10^{-5}$~\cite{baryo}, where the third uncertainty arises from uncertainty on the branching fraction of $\Lambda_c^+\rightarrow pK^-\pi^+$. This is the first measurement of this channel, with a significance above seven standard deviations.

%%%%%%%%%%%%%%%%%%%%%%%%%%%%%%%%%%%%%%%%%%%%%%%%%%%%%%%%%%%%%%%%%%%%%%%%%%%%%%%%%%%%%%%%%%%%%%%%%%%%%%%%%%%%%
%%%%%%%%%%%%%%%%%%%%%%%%%%%%%%%%%%%%%%%%%%%%%%%%%%%%%%%%%%%%%%%%%%%%%%%%%%%%%%%%%%%%%%%%%%%%%%%%%%%%%%%%%%%%%

\section{Charmless decays $B\rightarrow\eta h\ (h=\pi,K)$}\label{chapEtah}
Charmless decays are sensitive probes for the measurement of the CP violation. In the Standard Model, the decays $B\rightarrow\eta K$ proceed through $b\rightarrow s$ penguin and $b\rightarrow u$ tree transitions. The interference of these transitions can result in a large direct CP asymmetry $A_{CP}$~\cite{Lipkin}, defined as:
\begin{equation}
A_{CP}=\frac{\Gamma(\bar{B}\rightarrow\eta h)-\Gamma(B\rightarrow\eta\bar{h})}{\Gamma(\bar{B}\rightarrow\eta h)+\Gamma(B\rightarrow\eta\bar{h})},
\end{equation}
where $\Gamma(B\rightarrow\eta h)$ is the partial width obtained for the $B\rightarrow\eta h$ decay. Similar non-zero direct CP violation could be observed for $B^+\rightarrow\eta\pi^+$, given to the interference between $b\rightarrow d$ penguin and $b\rightarrow u$ tree diagrams.
Previous measurements by Belle~\cite{etahBelle} and BaBar~\cite{etahBabar} pointed to large negative $A_{CP}$, but preciser measurements are necessary to exclude the non-zero $A_{CP}$ in $B^+\rightarrow\eta\pi^+$. The branching fractions and $A_{CP}$ (for the charged modes) has been measured in the final Belle data sample~\cite{etah}, thus $772\times 10^6 B\bar{B}$, and are given in the Table~\ref{tab:etahBFACP}.
The first observation of $B^0\rightarrow\eta K^0$ is also reported, with a significance of $5.4\sigma$~\cite{etah}.

\begin{table}[htb]
  \caption{\label{tab:etahBFACP}
    Measured branching fractions $BF$ and direct CP asymmetry $A_{CP}$ of $B\rightarrow\eta h,\ h=K,\pi$.
    The first uncertainty is statistical, the second is systematic.}
  \begin{center}
    \begin{tabular}{lc}
      \hline\hline
      Observables & Measured values\\ \hline
      $BF(B^0\rightarrow\eta K^0)$ & $(1.27^{+0.33}_{-0.29} \pm 0.08)\times 10^{-6}$\\
      $BF(B^+\rightarrow\eta K^+)$ & $(2.12\pm 0.23 \pm 0.11)\times 10^{-6}$\\
      $BF(B^+\rightarrow\eta\pi^+)$ & $(4.07\pm 0.26 \pm 0.21)\times 10^{-6}$\\
      $A_{CP}(B^+\rightarrow\eta K^+)$ & $-0.38 \pm 0.11 \pm 0.01$ \\
      $A_{CP}(B^+\rightarrow\eta\pi^+)$ & $-0.19 \pm 0.06 \pm 0.01$\\
\hline\hline
\end{tabular}
\end{center}
\end{table}

%%%%%%%%%%%%%%%%%%%%%%%%%%%%%%%%%%%%%%%%%%%%%%%%%%%%%%%%%%%%%%%%%%%%%%%%%%%%%%%%%%%%%%%%%%%%%%%%%%%%%%%%%%%%%
%%%%%%%%%%%%%%%%%%%%%%%%%%%%%%%%%%%%%%%%%%%%%%%%%%%%%%%%%%%%%%%%%%%%%%%%%%%%%%%%%%%%%%%%%%%%%%%%%%%%%%%%%%%%%

\section{Charmless decays $B\rightarrow K\pi$}

In a similar way as for the $B\rightarrow\eta h$ decays (see Section~\ref{chapEtah}), the $B\rightarrow K\pi$ channels proceed through two diagrams: $b\rightarrow u$ tree and $b\rightarrow s$ penguins ones, both color-allowed or color-suppressed~\cite{nature}, whose interference are predicted to lead to a non-null direct CP assymetry $A_{CP}(K^{\pm}\pi^{\mp})$:
\begin{equation}
A_{CP}(K^{\pm}\pi^{\mp})=\frac{\Gamma(\bar{B}^0\rightarrow K^-\pi^+)-\Gamma(B^0\rightarrow K^+\pi^-)}{\Gamma(\bar{B}^0\rightarrow K^-\pi^+)+\Gamma(B^0\rightarrow K^+\pi^-)}.
\end{equation}
Previous measurements of the direct CP asymmetry in $B\rightarrow K\pi$ decays by Belle~\cite{nature} pointed a significant and unexplained difference between $A_{CP}(K^{\pm}\pi^{\mp})$ and $A_{CP}(K^{\pm}\pi^{0})$. Using the final sample, thus $772\times 10^6 B\bar{B}$ pairs plus an improved tracking, Belle measured the branching fractions and the direct asymmetries of $B\rightarrow K\pi$ modes~\cite{Kpi} (see Table~\ref{tab:KpiBFACP}). 
These values are compatible with the previous measurements by BaBar~\cite{Kpi_babar}, CDF~\cite{Kpi_CDF} and LHCb~\cite{Kpi_lhcb}. The possible isospin violating in $B\rightarrow K\pi$ decays can be investigated comparing the $BF$ ratios between the different modes with the SM prediction from the $SU(3)$ symmetry. The results, given in the Table~\ref{tab:KpiIso} are consistent with the different theoretical approaches~\cite{Kpi}.

\begin{table}[htb]
  \caption{\label{tab:KpiBFACP}
    Measured branching fractions $BF$ and direct CP asymmetry $A_{CP}$ of $B\rightarrow K\pi$.
    The first uncertainty is statistical, the second is systematic.}
  \begin{center}
    \begin{tabular}{lcc}
      \hline\hline
      Channel & $BF$ & $A_{CP}$\\ \hline
      $B^{\pm}\rightarrow K^{\pm}\pi^0$        & $(12.62 \pm 0.31 \pm 0.56)\times 10^{-6}$          & $0.043 \pm 0.024 \pm 0.002$ \\
      $B^0\rightarrow K^{\pm}\pi^{\mp}$        & $(20.00 \pm 0.34 \pm 0.63)\times 10^{-6}$          & $-0.069 \pm 0.014 \pm 0.007$ \\
      $B^{\pm}\rightarrow K^{0}\pi^{\pm}$      & $(23.97^{+0.53}_{-0.52} \pm 0.69)\times 10^{-6}$   & $-0.014 \pm 0.021 \pm 0.006$ \\
      $B^0\rightarrow K^0\pi^0$                & $(9.66 \pm 0.46 \pm 0.49) \times 10^{-6}$          & $-$ \\
\hline\hline
\end{tabular}
\end{center}
\end{table}

\begin{table}[htb]
  \caption{\label{tab:KpiIso}
    Widths $\Gamma$ ratios derived from the measured branching fractions (see Table~\ref{tab:KpiBFACP}), compared to the SM prediction from the $SU(3)$ symmetry.
    The first uncertainty is statistical, the second is systematic.}
  \begin{center}
    \begin{tabular}{lcc}
      \hline\hline
      Ratio & This measurement & $SM$\\ \hline
      $2\Gamma(K^+\pi^0)/\Gamma(K^0\pi^+)$ & $1.05 \pm 0.03 \pm 0.05$ & $1.15 \pm 0.05$\\
      $\Gamma(K^+\pi^-)/2\Gamma(K^0\pi^0)$ & $1.04 \pm 0.05 \pm 0.06$ & $1.12 \pm 0.05$\\
\hline\hline
\end{tabular}
\end{center}
\end{table}

%%%%%%%%%%%%%%%%%%%%%%%%%%%%%%%%%%%%%%%%%%%%%%%%%%%%%%%%%%%%%%%%%%%%%%%%%%%%%%%%%%%%%%%%%%%%%%%%%%%%%%%%%%%%%
%%%%%%%%%%%%%%%%%%%%%%%%%%%%%%%%%%%%%%%%%%%%%%%%%%%%%%%%%%%%%%%%%%%%%%%%%%%%%%%%%%%%%%%%%%%%%%%%%%%%%%%%%%%%%

\section{Observations of $B_{s}^0\rightarrow J/\psi f_0$ and $B_{s}^0\rightarrow J/\psi\eta$}
\begin{figure}[htb]
\begin{center}
\includegraphics[width=8cm]{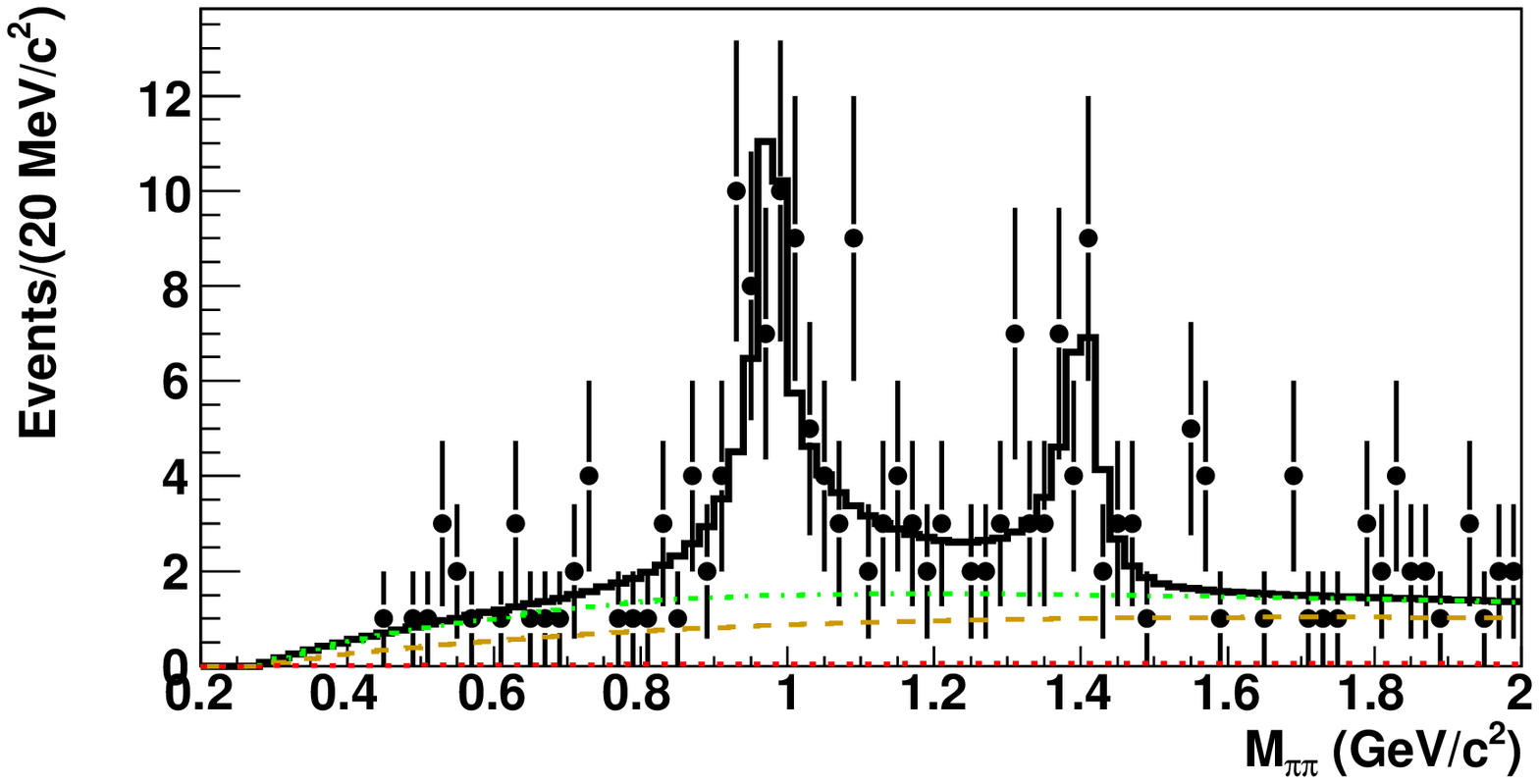}
\caption{Invariant mass of the di-pion system in data (points). The total fitted distribution is given by the solid line, the dash-dotted cuvred give the total background, the dashed curves other $J/\psi$ background, and the dotted curves show the non-resonant component. }
\label{fig:mpipi}
\end{center}
\end{figure}

The $b\rightarrow c\bar{c}s$ transition, occuring for instance in the decay $B^0_{s}\rightarrow J/\psi\phi$, benefits from a relatively large branching fraction. It has thus been used to extract the $B^0_s$ decay width difference $\Delta\Gamma$ and the CP violating phase $\beta_s$~\cite{jpsif0_CDF}\cite{jpsif0_D0}, sensitive to potential New Physics. 
Such study requires however an angular analysis, owing to the {\it Scalar$\rightarrow$Vector Vector} nature of the channel. The same $b\rightarrow c\bar{c}s$ transition can lead to the decay channel $B_{s}^0\rightarrow J/\psi f_0$, thus {\it Scalar$\rightarrow$Vector Scalar}, for which no angular analysis is so needed; furthermore leading order QCD, together with measurements of $D_s$ decays to $\phi$ and $f_0$ mesons, predicts its branching fraction to be $(3.1\pm 2.4)\times 10^{-4}$~\cite{jpsif0}. Using its final data sample at $\Upsilon(5S)$, thus $121.4/fb$ or $(1.24\pm 0.23)\times 10^7~B_s^*\bar{B}_s^*$ pairs, Belle measured the $B_{s}^0\rightarrow J/\psi f_0$ branching fraction, yielding together with LHCb~\cite{jpsif0_LHCb} its first observation~\cite{jpsif0}. The distributions of the invariant mass of the di-pion system from $f_0\rightarrow \pi^+\pi^-$ are given in the Figure~\ref{fig:mpipi}, where the $f_0(980)$ resonance can be seen, close to another scalar resonance, whose fitted parameters are: $m_0=(1.405\pm 0.015(\textrm{stat.})^{+0.001}_{-0.007}(\textrm{syst.}))~$GeV$/c^2$ and $\Gamma_0=(0.054\pm 0.033(\textrm{stat.})^{+0.014}_{-0.003}(\textrm{syst.}))$ GeV, which are consistent with the $f_0(1370)$ parameters~\cite{PDG}. The measured branching fractions, signal yields and significances are given in the Table~\ref{tab:jpsif0_res}.

\begin{table}[htb]
  \caption{\label{tab:jpsif0_res}
   Branching fractions, fitted signal yields and significance $S$ of the measurements performed in data on the $B_{s}^0\rightarrow J/\psi f_0(X)$ channels. The quoted uncertainties account for respectively the statistics, systematics and the number of $B_s^{(*)}\bar{B}_s^{(*)}$ in the data sample.}
  \begin{center}
    \begin{tabular}{lccc}
      \hline\hline
      Mode & Yield & $S$ & $BF\times 10^{-4}$\\ \hline
      $B_{s}^0\rightarrow J/\psi f_0(980)$ & $63^{+16}_{-10}$ & $8.4\sigma$ & $1.16^{+0.31 + 0.15 + 0.26}_{-0.19 -0.17 -0.18}$ \\
      $B_{s}^0\rightarrow J/\psi f_0(1370)$ & $19^{+6}_{-8}$     & $4.2\sigma$ & $0.34^{+0.11 +0.03 +0.08}_{-0.14 -0.02 -0.05}$ \\
\hline\hline
\end{tabular}
\end{center}
\end{table}

\begin{figure}[htb]
\begin{center}
\includegraphics[angle=-90,width=8cm]{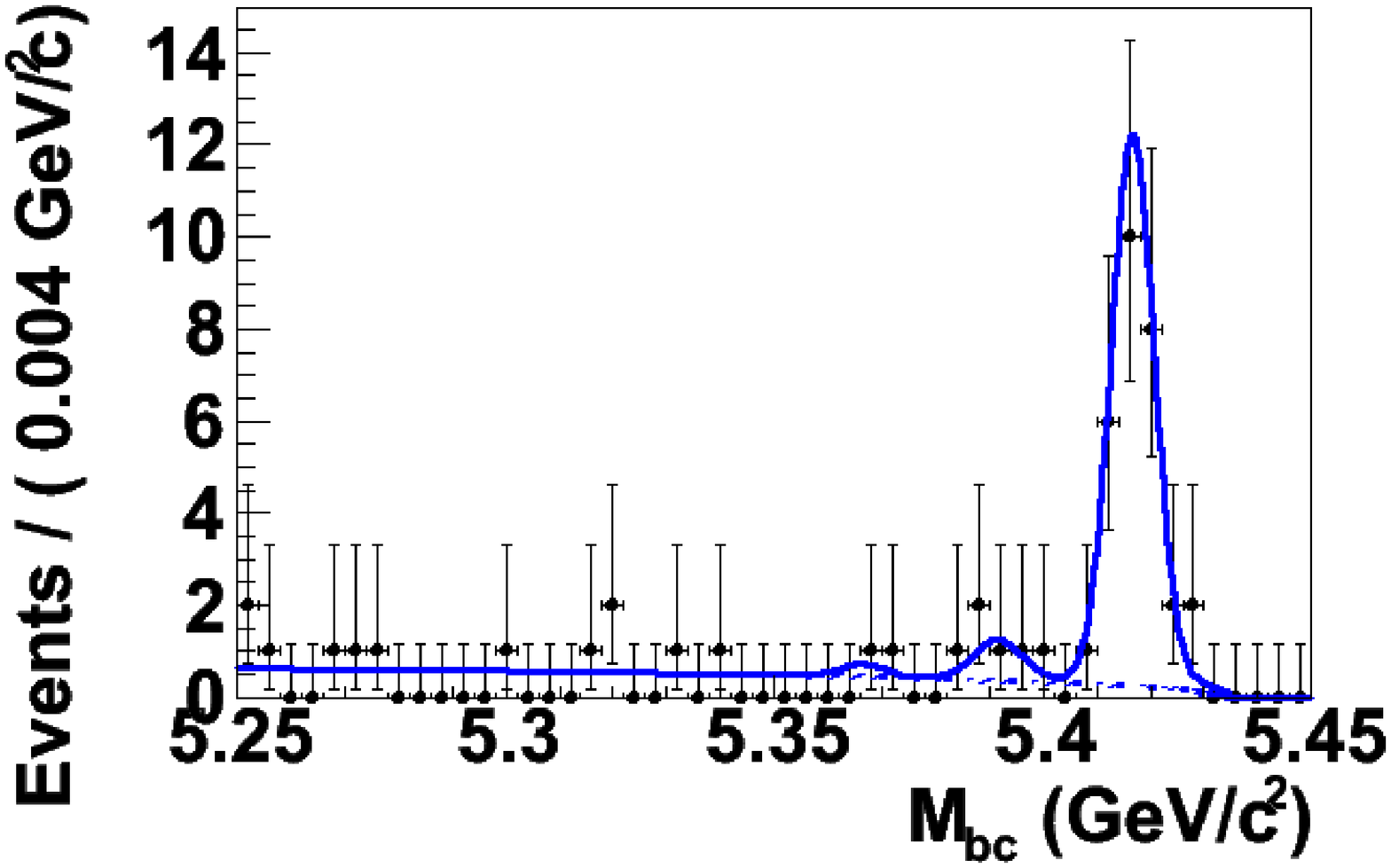}\\
\includegraphics[angle=-90,width=8cm]{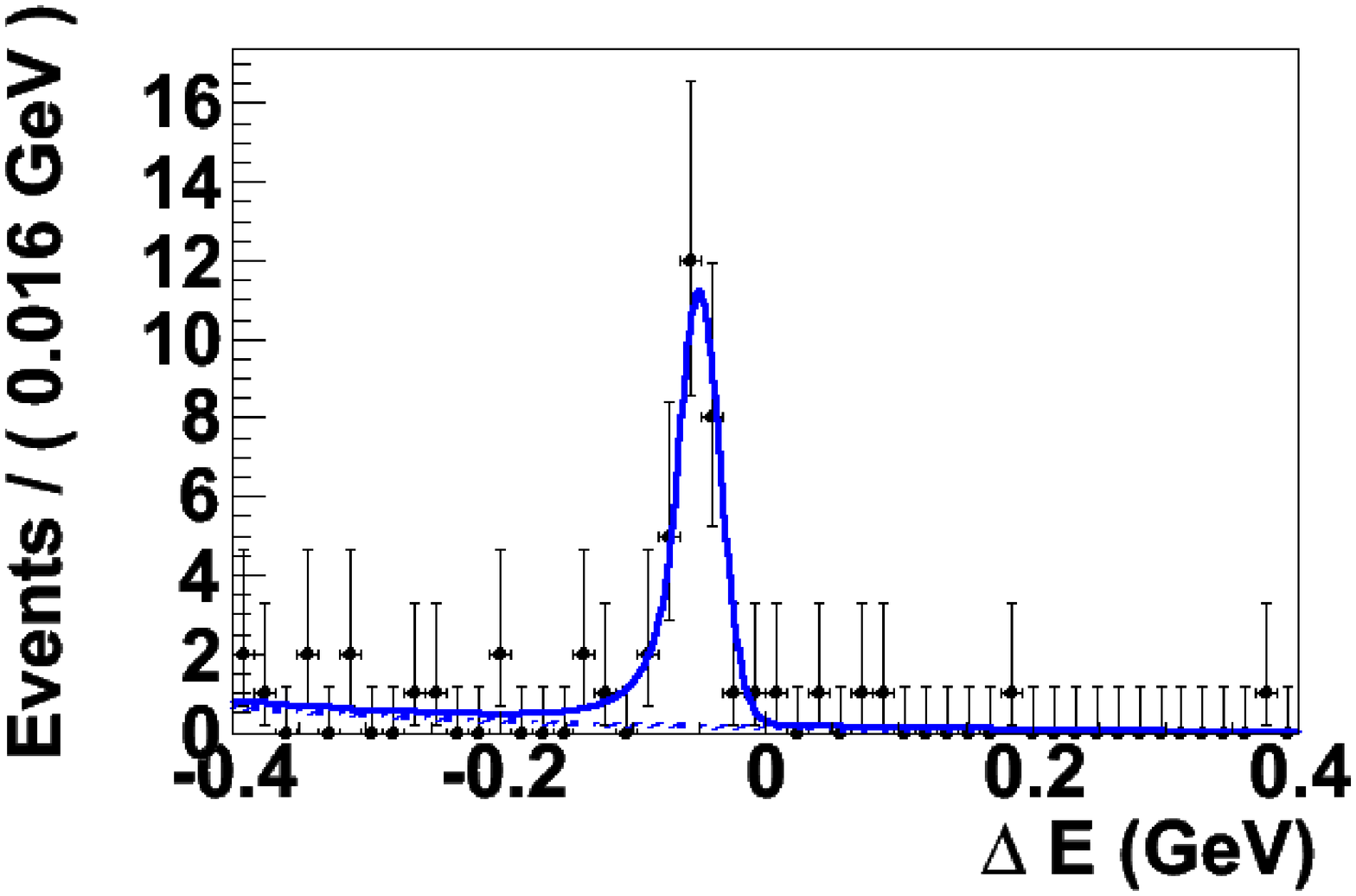}
\caption{The distributions in data (points) of the beam-constrained mass $M_{bc}$ and of the energy difference $\Delta E$~ for the sub-channel $B_s^0\rightarrow J/\psi\eta$ with $\eta\rightarrow\pi^+\pi^-\pi^0$. The total fit function is given by the solid line, the total background contribution by the dotted line, and the continuum background is represented by the dashed line.}\label{fig:mjpsieta}
\end{center}
\end{figure}

Belle also observed for the first time the decay $B_s^0\rightarrow J/\psi\eta$ using its full $\Upsilon(5S)$ dataset~\cite{jpsieta}. The distributions in data of the beam-constrained mass $M_{bc}$ and of the energy difference $\Delta E$~\cite{jpsif0} for the sub-channel $B_s^0\rightarrow J/\psi\eta$ with $\eta\rightarrow\pi^+\pi^-\pi^0$ are given in the Figure~\ref{fig:mjpsieta} where the $B$ signal can clearly be seen at $M_{bc}\simeq 5.42$ GeV$/c^2$ and $\Delta E\simeq 0$ GeV. The measured branching fraction yields:
\begin{equation}
BF(B_s^0\rightarrow J/\psi\eta) = (5.11 \pm 0.50 (\textrm{stat.}) \pm 0.35 (\textrm{syst.}) \pm 0.68 ({\textrm f_s})\times 10^{-4}),
\end{equation}
where the last uncertainty accounts for the $B_s^{(*)}\bar{B}_s^{(*)}$ production fraction at the $\Upsilon(5S)$.

The observation of these channels offers new CP channels for the study of the $B_s$ mixing property, paving the way for LHC experiments.

\bibliography{biblio}

\end{document}